\documentclass[conference,letterpaper]{IEEEtran}

\usepackage[utf8]{inputenc} 
\usepackage[T1]{fontenc}
\usepackage{url}
\usepackage{ifthen}
\usepackage{cite}
\usepackage[cmex10]{amsmath} 
\usepackage{mypreamble} 

\begin{document}
\title{A Provably Convergent Information Bottleneck Solution via ADMM} 


\author{%
  \IEEEauthorblockN{Teng-Hui Huang}
  \IEEEauthorblockA{Purdue University\\
                    Department of Electrical and Computer Engineering\\
                    West Lafayette, IN, USA\\
                    Email: huan1456@purdue.edu}
  \and
  \IEEEauthorblockN{Aly El Gamal}
  \IEEEauthorblockA{Purdue University\\
                    Department of Electrical and Computer Engineering\\
                    West Lafayette, IN, USA\\
                    Email: 
                    elgamala@purdue.edu}
}


\maketitle

\begin{abstract}
The Information bottleneck (IB) method enables optimizing over the trade-off between compression of data and prediction accuracy of learned representations, and has successfully and robustly been applied to both supervised and unsupervised representation learning problems. However, IB has several limitations. First, the IB problem is hard to optimize. The IB Lagrangian $\mathcal{L}_{IB}:=I(X;Z)-\beta I(Y;Z)$ is non-convex and existing solutions guarantee only local convergence. As a result, the obtained solutions depend on initialization. Second, the evaluation of a solution is also a challenging task. Conventionally, it resorts to characterizing the information plane, that is, plotting $I(Y;Z)$ versus $I(X;Z)$ for all solutions obtained from different initial points. 
Furthermore, the IB Lagrangian has phase transitions while varying the multiplier $\beta$. At phase transitions, both $I(X;Z)$ and $I(Y;Z)$ increase abruptly and the rate of convergence becomes significantly slow for existing solutions. Recent works with IB adopt variational surrogate bounds to the IB Lagrangian. Although allowing efficient optimization, how close are these surrogates to the IB Lagrangian is not clear.
In this work, we solve the IB Lagrangian using augmented Lagrangian methods. With augmented variables, we show that the IB objective can be solved with the alternating direction method of multipliers (ADMM). Different from prior works, we prove that the proposed algorithm is consistently convergent, regardless of the value of $\beta$. Empirically, our gradient-descent-based method results in information plane points that are comparable to those obtained through the conventional Blahut-Arimoto-based solvers and is convergent for a wider range of the penalty coefficient than previous ADMM solvers.
\end{abstract}


\section{Introduction}\label{sec:intro}

The Information bottleneck (IB) formulation\cite{tishby2000information} aligns with the goals of important problems such as feature extraction and representation learning, and therefore the literature in various research fields have adopted IB and shown fruitful results\cite{Kolchinsky_2019,DBLP:journals/corr/AlemiFD016,makhdoumi2014information,4914747}. Given the joint probability of observations $X$ and latent features $Y$, IB methods aim at finding a representation of raw observations that is minimal in expression complexity but satisfying relevance criterion toward its meaning. The IB 
objective is formulated as a constrained optimization problem\cite{tishby2000information}:
\begin{equation}
    \begin{split}
        \text{minimize}_{p(z|x)}&\quad I(X;Z)\\
        \text{subject to}: &\quad I(Y;Z)>I_0,\\
        &\sum_{z}p(z|x) = 1, \forall x\in\mathcal{X},\\
        & Y-X-Z \quad \text{Markov chain},
    \end{split}
\end{equation}
where the involved random variables are as follows: $X$ represents the observations, $Y$ the relevant target and $Z$ the latent representation. The constant threshold $I_0$ controls the trade-off between $I(X;Z)$ and $I(Y;Z)$. The IB problem can be solved with the Lagrangian multiplier approach:
\begin{equation}\label{eq:IB_lag}
\mathcal{L}_{IB}:=I(X;Z)-\beta I(Y;Z)+\sum_{x}\lambda_x \left(\sum_{z}p(z|x)-1\right),
\end{equation}
where the Lagrangian multiplier $\beta$ is a fixed constant when employed in a Blahut-Arimoto (BA) solver\cite{1054855}, $\{\lambda_i\}$ are other multipliers for equality constraints, imposed to ensure valid conditional probability. Using the first order functional derivative w.r.t. $p(z|x),\forall z\in\mathcal{Z},\forall x\in\mathcal{X}$, one can obtain the well known BA-based algorithm \cite{1054855}:
\begin{equation}\label{eq:ib_selfconsist}
p^{k+1}(z|x)=\frac{p^{k}(z)}{\mathcal{K}(x,\beta)}\exp\big\{-\beta D_{KL}[p(y|x)||p^{k}(y|z)]\big\},
\end{equation}
where $D_{KL}(\mu||\nu)$ is the Kullback-Leibler divergence, $\mathcal{K}(x,\beta)=\sum_{z'}p^k(z')\exp\{-\beta D_{KL}[p(y|x)||p^k(y|z')]\}$ is the normalization function and the superscript is the iteration counter. Then, $p^{k+1}(z):=\sum_{x}p^{k+1}(z|x)p(x)$ and $p^{k+1}(z|y):=\sum_{x}p^{k+1}(z|x)p(x|y)$ are updated accordingly. However, since \eqref{eq:IB_lag} lacks convexity, a solution from the BA-based algorithm is not guaranteed to be globally optimal\cite{tishby2000information,7996419}. The evaluation of the performance of the solutions therefore requires sweeping through a range of the trade-off parameter $\beta$ and plotting the obtained pairs of $I(X;Z),I(Y;Z)$ in $x,y$ axes. This plot is called the information plane\cite{tishby2000information,slonim2000agglomerative}. Generally, $I(Y;Z)$ that is higher for a fixed $I(X;Z)$ is considered as a better solution. The characterization of the information plane, in turn, is a challenging task. Although following the data processing inequality $I(X;Z)\geq I(Y;Z)$, the optimal solutions to \eqref{eq:IB_lag} are trivial for $\beta\leq 1$\cite{IBlearnability2019,Wu2020Phase}, at certain $\beta>1$, corresponding to what are known as the IB phase transitions \cite{Wu2020Phase,Gedeon2012,5420292}, the iterations for convergence will increase dramatically.

A recent work \cite{8919752} adopted the alternating direction method of multipliers (ADMM\cite{boyd2011distributed}), and a relaxation of the well-studied augmented Lagrangian methods (ALM)\cite{BertsekasDimitriP.1999Np}, to solve the IB Lagrangian. ALM can asymptotically lead to optimal solutions as in the original Lagrangian and is simple to implement with iterative solvers\cite{NocedalJorge2006No}. 
The focus of \cite{8919752} was on addressing the slow convergence of the BA-based algorithm in solving IB but there is no proof of convergence 
for the algorithm proposed therein. 

Inspired by this idea, we introduce a simpler IB solver with ADMM. Compared to \cite{8919752}, our method has fewer augmented variables and less constraints. Most importantly, we prove that our algorithm is convergent, based on the recent non-convex, non-smooth convergence results of ADMM\cite{wang2014convergence,zhang_shen_2019}. In proving convergence, we relax the traditionally needed Lipschitz differentiability assumption by exploiting the semi-convexity of the decoupled objective function. We further show that the 
weak convexity assumption in \cite{zhang_shen_2019} is not required in our case. Empirically, we implement  gradient-based methods for both the proposed ADMM solver and the IB method. We find our method to consistently converge and deliver denser points on the information-plane comparable to those obtained from the BA-based algorithm. Moreover, our method is convergent for a significantly wider range of the penalty coefficient than \cite{8919752}.

\section{Problem Formulation}\label{sec:II_problemform}
We introduce $\{p(z)\}$ as a set of augmented variables in addition to $\{p(z|x)\}$. Besides the resulting simplicity as will be clear in the sequel, this allows us to decompose the IB Lagrangian into two separate sub-problems. In the following, we represent \eqref{eq:IB_lag} as ALM. Define the dual multipliers $\mu\in\mathcal{U},|\mathcal{U}|=|\mathcal{Z}|$ and the corresponding penalty $\{p(z_i)-\sum_{x}p(z_i|x)p(x)\}_{
i=1}^{|\mathcal{Z}|}$. To simplify notation, let $N_x$ denote the cardinality of a random variable $X$, e.g. $N_x=|\mathcal{X}|$, we define the vectors:
\begin{equation*}
\begin{split}
    p_z:=&[p(z_1),\cdots,p(z_{N_z})]^T,\\
    p_{z|x}:=&\begin{bmatrix}{p(z_1|x_1),...,p(z_1|x_{N_x}),p(z_2|x_1)...,p(z_{N_z}|x_{N_x})}\end{bmatrix}^T,\\
    \mu_z:=&[\mu(1),\cdots,\mu(N_z)]^T,
\end{split}
\end{equation*}
where we cascade the conditional probability $p(z|x),\forall z\in\mathcal{Z},\forall x\in\mathcal{X}$ into a long vector $p_{z|x}$. Then, we can transform the IB Lagrangian into standard ALM form:
\begin{equation}\label{eq:ib_alm}
    \begin{split}
        \mathcal{L}_{c}(p_z,p_{z|x},\mu)&=I(X;Z)-\beta I(Y;Z)\\
        &\quad+\lambda_x^T\bigl(Jp_{z|x}-\mathbf{1}_{N_x}\bigr)+\lambda_z\bigl(\mathbf{1}_{N_z}^Tp_z-1\bigr)\\
        &\quad+{\mu_z^T\bigl(p_z-Bp_{z|x}}\bigr)\\
        &\quad+\frac{c}{2}\Norm{p_z-Bp_{z|x}}^2.
    \end{split}
\end{equation}
In above, $\mathbf{1}_k\in\mathbb{R}^k$ denotes the all-one vector. $I_k$ is the identity matrix with $k$ ones on the main diagonal and zeros elsewhere, $B:=I_{N_z}\otimes p_x^T$ with $\otimes$ denoting the Kronecker product and $J:=\mathbf{1}_{N_z}^T\otimes I_{N_x}$. Note that the two terms in the second line of \eqref{eq:ib_alm} capture equality constraints for the primal variables to be valid probabilities with multipliers $\lambda_z\in\mathbb{R}$ and $\lambda_{x}\in\mathbb{R}^{N_x}$. Finally, the last line is the quadratic penalty to reinforce $\sum_{x}p(x)p(z|x)=p(z)$ with a fixed scalar coefficient $c>0$ and 
$\norm{\cdot}$ is the $2$-norm if not explicitly specified in the subscript.

Compared to the IB Lagrangian in \eqref{eq:IB_lag}, we treat $\{p(z)\}$ as parameters and impose penalty to encourage in the $k^{\text{th}}$ step of the iterative solution that $p^k(z)=\sum_{x}p^k{(z|x)p(x)}$, while in the BA algorithm, $\{p_{BA}(z)\}$ is forced to be the marginal probability, i.e., in each iteration $k$, $p_{BA}^k(z)=\sum_{x}p_{BA}^k(z|x)p(x)$. In this sense, we relax the equality constraint of BA, exploit the created structure and apply ADMM\cite{boyd2011distributed} to solve the ALM efficiently. This is due to the fact that when the objective function of ALM can be decomposed into separate sub-problems w.r.t. its primal variables, it can be solved in an alternating fashion\cite{BertsekasDimitriP.1999Np}. Indeed, it is straightforward to see that: 
\begin{align*}
  I(X;Z)-\beta I(Y;Z)&=\underbrace{(1-\beta)H(Z)}_{F_\beta(p_z)}\\
  \quad&+\underbrace{\beta H(Z|Y)-H(Z|X)}_{G_\beta(p_{z|x})},
\end{align*}
where we decompose $I(X;Z),I(Y;Z)$ into  $F_\beta(p_{z})$ and $G_\beta(p_{z|x})$ w.r.t. $p_z$ and $p_{z|x}$ respectively:
\begin{equation}\label{eq:fg_def}
\begin{split}
F_\beta(p_z)&:=(\beta-1)\Pthes{\sum_{i=1}^{N_z}p(z_i)\log{p(z_i)}},\\
G_\beta(p_{z|x})&:=\begin{multlined}[t] \sum_{i=1}^{N_x}p(x_i)\sum_{j=1}^{N_z}p(z_j|x_i)\log{p(z_j|x_i)}\\
-\beta\sum_{l=1}^{N_y}p(y_l)\sum_{m=1}^{N_z}\Pthes{\sum_{r=1}^{N_x}p(z_m|x_r)p(x_r|y_l)}\\
\cdot\log{\Pthes{\sum_{r=1}^{N_x}p(z_m|x_r)p(x_r|y_l)}}.
\end{multlined}
\end{split}
\end{equation}
Note that in $G_\beta(p_\mathbf{z|x})$ we use the Markov chain condition: $Y-X-Z$, that is, $p(z|y)=\sum_{x}p(z|x)p(x|y)$. For $\beta>1$, $F_\beta$ is the negative entropy function and therefore is convex w.r.t. $p_z$. On the other hand, $G_\beta$ is related to the conditional entropy of $p_{z|x}$ but not convex. This aligns with the non-convexity of the IB Lagrangian \cite{tishby2000information}.
In contrast to \cite{8919752}, our method is easier to optimize. \cite{8919752} treated both $\{p_z\}$ and $\{p_{z|y}\}$ as augmented variables, while we only add $\{p_z\}$. Together with $\{p_{z|x}\}$, they have $N_y\times N_z$ extra variables and consequently impose $N_y$ more penalty terms. Empirically, their algorithm diverges even with a high penalty coefficient $c$, whereas ours is convergent 
even with a relatively low $c$.
\section{Proposed Approach}\label{sec:III_proposedApp}
\subsection{Augmented IB Lagrangian}\label{para:main_algorithm}
The proposed IB solution relies on ADMM by optimizing over $p_{z|x},p_z$ and updating the dual multiplier $\mu_z$ alternately:
\begin{equation}\label{eq:main_alg}
    \begin{split}
        p^{k+1}_{z|x}=&\underset{Jp_{z|x}=\mathbf{1}_{N_z}}{\arg\min}\mathcal{L}_{c}(p^{k}_z,p_{z|x},\mu_z^k),\\
        p^{k+1}_z=&\underset{\mathbf{1}_{N_z}^Tp_z=1}{\arg\min}\quad\mathcal{L}_{c}(p_z,p^{k+1}_{z|x},\mu_z^k)+\omega D_\phi(p_z||p_z^k),\\
        \mu_z^{k+1} =&\mu_z^k + c(p^{k+1}_z-Bp^{k+1}_{z|x}),
    \end{split}
\end{equation}
where $\omega\geq 0$ is a scalar coefficient.
Note that we introduce $D_\phi$ to regularize the consecutive updates of $p_z$. This follows from the Bregman ADMM\cite{wang2014bregman,wang2014convergence}.

\subsection{Regularization with Bregman Divergence}
A common approach in facilitating the optimization with the ADMM algorithm is combining Bregman divergence \cite{wang2014bregman}. When updating the primal variables, a properly chosen Bregman divergence function improves convergence.
A Bregman divergence $D_\phi(u||v)$ follows:
\begin{equation*}
    D_\phi(u||v):=\phi(u)-\phi(v)-\angdot{\nabla\phi(v),u-v},
\end{equation*}
where $\langle u,v \rangle$ denotes the inner product of $u,v\in \mathbb{R}^n$. In our case, since $F_\beta(p_z)$ is the negative entropy function, we choose $\phi(u):=\sum u\log(u)$ so that $D_\phi(u||v)$ becomes the KL-divergence. 
In proving convergence of the proposed algorithm, by increasing the coefficient $\omega$, the Bregman divergence in \eqref{eq:main_alg} strengthens the convexity of $F_\beta$, which regularizes consecutive updates of $p_z$; However, empirically, a large $\omega$ will render the convergence of the algorithm slow. We defer the details to the next section.

\section{Convergence Analysis}\label{sec:IVconverge}
Generally, the convergence of ADMM algorithms depends on the convexity of the separate components of the objective function, through which alternating iterative optimization is applied. Recently, the convergence of ADMM solvers was generalized to non-convex functions. While this line of research shows mostly local convergence, in some cases, global convergence is possible under mild assumptions. The results that are most relevant to our case are in \cite{zhang_shen_2019}. We also use \cite{wang2014convergence} and \cite{doi:10.1137/140998135} as references in our proof.
\subsection{Preliminaries}
In the literature, the convergence analyses of ADMM, for non-convex, non-smooth cases, rely on the assumption of the Lipschitz differentiability of the objective function w.r.t. primal variables and then the application of the Kurdyka-Łojasiewicz inequality\cite{Kurdyka1998}. For our case and the IB problem, however, such an assumption of uniform Lipschitz differentiability would not hold. This is straightforward to see from the functional derivative of $
F_\beta(p_z),G_\beta(p_{z|x})$ w.r.t. $p(z)$ and $p(z|x)$ respectively.
Recently, in \cite{zhang_shen_2019}, the authors showed that if one of the two decomposed objective functions is strongly and the other weakly convex, then the convergence is assured without both Lipschitz differentiability and the Kurdyka-Łojasiewicz inequality. In addition, we further show that the weak convexity condition can be relaxed.
Also interestingly, we show that the proposed algorithm is convergent independent of the positive penalty coefficient $c$.

\begin{definition}
A convex function $f(x)$ is $\eta$-\textit{strongly convex} if there exists a scalar $\eta>0$ such that $\langle{\nabla f(y)-\nabla f(x),y-x}\rangle\geq \eta\lVert{y-x}\rVert^2$. Or equivalently, $\nabla_x^2f(x)-\eta I$ is positive semi-definite.
\end{definition}
It is well known that the negative entropy function $F_\beta(p_z)$ is $(\beta-1)$-strongly convex in $1$-norm\cite{Cover:2006:EIT:1146355}, and consequently in $2$-norm. 
Before we present the main proof, we have the following lemma that justifies our claim that $D_{KL}(p_z^{k+1}||p_z^k)$ in the proposed algorithm strengthens the convexity of $F_\beta(p_z)$.
\begin{lemma}\label{lemma:bregman_str_cvx}
For the proposed algorithm, suppose there exists a stationary point $q^*:=(p_z^*,p_{z|x}^*,\mu_z^*)$, then $F_\beta(p_z)+\omega D_{KL}(p_z||p_z^{k-1})$ is $(\beta+\omega-1)$-strongly convex w.r.t. $p_z$ in 2-norm.
\end{lemma}
\begin{IEEEproof}
See Appendix \ref{app:pf_of_breg_reg}.
\end{IEEEproof}

\subsection{Proof of Convergence}

In proving the convergence of the proposed algorithm using Lemma \ref{lemma:bregman_str_cvx}, we can denote $F_\beta(p_z)+\omega D_{KL}(p_z||p_z^{k-1})$ as an $\eta_z$-\textit{strongly convex} function, namely, $\eta_z=\beta+\omega-1$.  Additionally, we make the following assumptions:
\begin{itemize}\label{parag:assumptions}
    \item There exists at least one stationary point $(p_z^*,p_{z|x}^*,\mu_z^*)$ of $\mathcal{L}_c(p_z,p_{z|x},\mu_z)$.
    \item $p(y|x) > 0,\forall x\in\mathcal{X},\forall y\in\mathcal{Y}$.
    \item $\exists \varepsilon>0$ such that $p(z|x)\geq \varepsilon, \forall x \in\mathcal{X},\forall z\in\mathcal{Z}$.
\end{itemize}
From the minimizer conditions and the steps of the proposed algorithm:
\begin{equation}\label{eq:min_condition}
    \begin{split}
        \nabla G_\beta(p^{k+1}_{z|x}) &= B^T\mu_z^{k}+cB^T\pthes{p_z^k-Bp_{z|x}^{k+1}}\\
        &=B^T\mu_z^{k+1}+cB^T\pthes{p_z^k-p_z^{k+1}},\\
        \nabla F_\beta(p_z^{k+1})&=
        \begin{multlined}[t]
        -\mu^{k}-c\pthes{p_z^{k+1}-Bp_{z|x}^{k+1}}\\
        -\omega\left[\nabla\phi(p_z^{k+1})-\nabla\phi(p_z^k)\right]
        \end{multlined}\\
        &=-\mu_z^{k+1}-\omega\big[\nabla\phi(p_z^{k+1})-\nabla\phi(p_z^k)\big],\\
        \mu_z^{k+1}&=\mu_z^k+c\pthes{p_z^{k+1}-Bp_{z|x}^{k+1}}.
    \end{split}
\end{equation}
Then it follows from the assumptions above, at a stationary point $q^*:=\{p_z^*,p_{z|x}^*,\mu_z^*\}$, we have:
\begin{equation}\label{eq:sta_pts}
    \begin{split}
        p_z^*&=Bp_{z|x}^*,\\
        \nabla_z\mathcal{L}_c^*&=\nabla_zF_\beta^*+\mu_z^*=0,\\
        \nabla_{z|x}\mathcal{L}_c^*&=\nabla_{z|x}G_\beta^*-B^T\mu_z^*=0.
    \end{split}
\end{equation}
Following \cite{zhang_shen_2019} and \cite{boyd2011distributed}, we exploit the strong convexity of $F_\beta(p_z)$ and obtain a Lyapunov function that guarantees convergence to a stationary point asymptotically. In contrast to the assumptions made in the reference works, here $G_\beta(p_{z|x})$ is not required to be weakly convex. To simplify notation, we define $\Delta_* r^{k}:=r^{k}-r^{*}$, which indicates the difference in value at the $k^{\text{th}}$ step from a stationary point of the corresponding variable $r$, and $E_{r}[\cdot]$ denotes the expectation w.r.t. $r$.
\begin{lemma}\label{lemma:gzx_coeff}
With $G_\beta(p_{z|x})$ defined as in \eqref{eq:fg_def} and the assumption $p(z|x)\geq \varepsilon>0. \forall x\in\mathcal{X},z\in\mathcal{Z}$, we have:
\begin{equation}\label{eq:g_grad}
    \begin{split}
        \angdot{\Delta_*\nabla G_\beta(p_{z|x}^{k+1}),\Delta_* p_{z|x}^{k+1}}&\geq -\gamma_\beta \norm{B\Delta_*p^{k+1}_{z|x}}^2,
    \end{split}
\end{equation}
where $\gamma_\beta:=\beta\kappa/\varepsilon-1$ and  $\kappa=\sup_y\left(\frac{\max_{x\in\mathcal{X}}p(y|x)}{\min_{x'\in\mathcal{X}}p(y|x')}-1\right)^2$.
\end{lemma}
\begin{IEEEproof}
Summing all $N_x\times N_z$ terms of the functional derivative of $G_\beta(p_{z|x})$ w.r.t. $p_{z|x}$, in the left of \eqref{eq:g_grad}, we have:
\begin{multline*}
    E_{x}\left[D_{KL}(p_{z|X}^{k+1}||p_{z|X}^*)\right]+E_{x}\left[D_{KL}(p_{z|X}^*||p_{z|X}^{k+1})\right]\\
    -\beta \left\{E_{y}\left[D_{KL}(p_{z|Y}^{k+1}||p_{z|Y}^*)\right]+E_{y}\left[D_{KL}(p_{z|Y}^*||p_{z|Y}^{k+1})\right]\right\},
\end{multline*}
where the upper case letter in the subscript denotes a random variable. For the first term above, applying the log-sum inequality \cite{Cover:2006:EIT:1146355} and then Pinsker's inequality sequentially, we have:
\begin{equation*}
    \begin{split}
        E_x[D_{KL}(p_{z|X}^{k+1}||p_{z|X}^*)]&\geq D_{KL}(E_{x}[p_{z|X}^{k+1}]||E_{x}[p_{z|X}^*])\\
        &\geq \frac{1}{2}\norm{B\Delta_* p_{z|x}^{k+1}}_1^2\\
        &\geq \frac{1}{2}\norm{B\Delta_* p_{z|x}^{k+1}}_2^2.
    \end{split}
\end{equation*}
Similarly, the inequality holds for the second term. 
For the terms related to $p(z|y)$, note that there is a negative sign before $\beta$ so we need to upper bound $D_{KL}[p_{z|y}^{k+1}||p_{z|y}^*]$ for each given $y$. Following the method in \cite{6686179}, rewrite $p_{z|y}:=E_{x|y}[p_{z|X}]$ and for each $y\in\mathcal{Y}$:
\begin{equation}\label{eq:pv_ineq_abs}
    \begin{split}
        {}&D_{KL}\pthes{p_{z|y}^{k+1}||p_{z|y}^*}+D_{KL}\pthes{p_{z|y}^*||p_{z|y}^{k+1}}\\
        =&\sum_{z}\pthes{E_{x|y}[p_{z|X}^{k+1}]-E_{x|y}[p_{z|X}^*]}\log{\frac{E_{x|y}[p_{z|X}^{k+1}]}{E_{x|y}[p_{z|X}^{*}]}}\\
        \leq& \sum_{z}\bigg|E_{x|y}[p_{z|X}^{k+1}]-E_{x|y}[p_{z|X}^*]\bigg|\bigg|\log{\frac{E_{x|y}[p_{z|X}^{k+1}]}{E_{x|y}[p_{z|X}^*]}}\bigg|.
    \end{split}
\end{equation}
We then bound the two absolute terms in \eqref{eq:pv_ineq_abs} separately. For the second term in \eqref{eq:pv_ineq_abs}, considering the following identity and the fact that $\log{x}<x-1$ for $x>0$, we have:
\begin{equation}\label{eq:log_ineq}
    \begin{cases}
        a>b, & \log{\frac{a}{b}}\leq\frac{a}{b}-1=\frac{a-b}{b}\\
        b>a, & \log{\frac{b}{a}}\leq\frac{b}{a}-1=\frac{b-a}{a}
    \end{cases}
    \Rightarrow
    \left|\log{\frac{a}{b}}\right|\leq \frac{\left|a-b\right|}{\min\{a,b\}}.
\end{equation}
Let $a=E_{x|y}[p_{z|X}^{k+1}],b=E_{x|y}[p_{z|X}^*]$, we obtain an upper bound. For the first term in \eqref{eq:pv_ineq_abs}, we use the following relation:
\begin{equation}\label{eq:excy_diff}
    \begin{split}
        &E_{x|y}[p_{z|X}^{k+1}]-E_{x|y}[p_{z|X}^*]\\
        {}=&\sum_{x}p^{k+1}(z|x)p(x)\frac{p(y|x)}{p(y)}-\sum_{x}p^*(z|x)p(x)\frac{p(y|x)}{p(y)}\\
        {}=&\sum_{x}\delta^{\geq}_{z|x}\left(k+1, *\right)\left[p^{k+1}(z|x)-p^*(z|x)\right]\frac{p(x)p(y|x)}{p(y)}\\
        &+\sum_{x}\delta^{<}_{z|x}\left(k+1,*\right)\left[p^{k+1}(z|x)-p^*(z|x)\right]\frac{p(x)p(y|x)}{p(y)}\\
        \leq& \sup_{x\in\mathcal{X}}\frac{p(y|x)}{p(y)}\sum_x\left[p^{k+1}(z|x)-p^*(z|x)\right]p(x)\\
        &-\inf_{x\in\mathcal{X}}\frac{p(y|x)}{p(y)}\sum_x\left[p^{k+1}(z|x)-p^*(z|x)\right]p(x)\\
        =&\left(\kappa_y-1\right)\inf_{x\in\mathcal{X}}\frac{p(y|x)}{p(y)}\left(E_{x}[p^{k+1}_{z|X}]-E_{x}[p^*_{z|X}]\right),
    \end{split}
\end{equation}
where $\delta^{\circ}_{z|x}(k,l)$ denotes an indicator function with an relation operator $\circ\in\{\geq,<\}$. The function equals to one if the argument $[p^k(z|x)\circ p^l(z|x)]$ is true and zero otherwise, and we define $\kappa_y:=\left(\frac{\sup_{x\in\mathcal{X}}p(y|x)}{\inf_{x'\in\mathcal{X}}p(y|x')}\right)$. Combining \eqref{eq:log_ineq} and \eqref{eq:excy_diff}, we have:
\begin{equation}\label{eq:single_dkl_pyz}
    \begin{split}
        {}&\begin{multlined}[t]
        D_{KL}\pthes{E_{x|y}[p_{z|X}^{k+1}]||E_{x|y}[p_{z|X}^{*}]}\\
            +D_{KL}\pthes{E_{x|y}[p_{z|X}^{*}]||E_{x|y}[p_{z|X}^{k+1}]}
        \end{multlined}\\
        \leq& \sum_{z}\bigg|E_{x|y}[p_{z|X}^{k+1}]-E_{x|y}[p_{z|X}^*]\bigg|\bigg|\log{\frac{E_{x|y}[p_{z|X}^{k+1}]}{E_{x|y}[p_{z|X}^*]}}\bigg|\\
        \leq&\sum_z\frac{\bigg|E_{x|y}[p_{z|X}^{k+1}]-E_{x|y}[p_{z|X}^{*}]\bigg|^2}{\min\{ E_{x|y}[p_{z|X}^{k+1}],E_{x|y}[p_{z|X}^*]\}}
        .
    \end{split}
\end{equation}
Now consider the following:
\begin{equation*}
E_{x|y}[p_{z|X}]=\sum_x\frac{p(z|x)p(y|x)p(x)}{p(y)}\geq \inf_{x\in\mathcal{X}}\frac{p(y|x)}{p(y)}p(z),
\end{equation*}
and under the assumption $p(z|x)\geq \varepsilon$ in \ref{parag:assumptions}, along with the Markov chain condition $Y-X-Z$:
\begin{align*}
p(z|y)&=\sum_{x}p(z|x)p(x|y)\\ &\geq\inf_{x\in\mathcal{X}}\frac{p(y|x)}{p(y)}\sum_{x\in\mathcal{X}}p(z|x)p(x)\geq \inf_{x\in\mathcal{X}}\frac{p(y|x)}{p(y)}\varepsilon.
\end{align*}
Then we have:
\begin{equation*}
    \begin{split}
        {}&\sum_z\frac{\left|E_{x|y}[p^{k+1}_{z|X}]-E_{x|y}[p^{*}_{z|X}]\right|^2}{\min_{}\{E_{x|y}[p^{k+1}_{z|X}],E_{x|y}[p^*_{z|X}]\}}\\
        \leq&\begin{multlined}[t]
            (\kappa_y-1)^2\sum_z\frac{\left[\inf_{x\in\mathcal{X}}p(y|x)/p(y)\right]^2}{\min_{}\{p^{k+1}(z|y),p^*(z|y)\}}\\
            \cdot\norm{\sum_{x}p^{k+1}(z|x)p(x)-\sum_xp(x)p^*(z|x)}^2
        \end{multlined}\\
        \leq& \begin{multlined}[t]
            (\kappa_y-1)^2\inf_{x\in\mathcal{X}}\frac{p(y|x)}{p(y)}\sum_z\frac{\inf_{x\in\mathcal{X}}p(y|x)/p(y)}{\min{\{p^{k+1}(z|y),p^*(z|y)\}}}\\
            \cdot\norm{E_x[p^{k+1}_{z|X}]-E_x[p^*_{z|X}]}^2
        \end{multlined}\\
        \leq&\frac{(\kappa_y-1)^2}{\varepsilon}\inf_{x\in\mathcal{X}}\frac{p(y|x)}{p(y)}\norm{B\Delta_*p_{z|x}^{k+1}}_2^2,
    \end{split}
\end{equation*}
Combining the above with \eqref{eq:single_dkl_pyz}, we have:
\begin{equation*}
    \begin{split}
        {}&E_y\left[D_{KL}(p_{z|Y}^{k+1}||p_{z|Y}^*)+D_{KL}(p_{z|Y}^*||p_{z|Y}^{k+1})\right]\\
        &\leq \sum_{y\in\mathcal{Y}}p(y)\frac{(\kappa_y-1)^2}{\varepsilon}\inf_{x\in\mathcal{X}}\frac{p(y|x)}{p(y)}\norm{B\Delta_*p_{z|x}^{k+1}}_2^2\\
        &=\frac{1}{\varepsilon}\sum_y\inf_{x\in\mathcal{X}}p(y|x)(\kappa_y-1)^2\norm{B\Delta_*p_{z|x}^{k+1}}_2^2.
    \end{split}
\end{equation*}
Finally, by defining $\kappa:=\underset{y}{\sup}(\kappa_y-1)^2$, we have:
\begin{equation*}
    \begin{split}
        \angdot{\Delta_*G_\beta(p_{z|x}^{k+1}),\Delta_* p_{z|x}^{k+1}}&\geq \norm{B\Delta_*p_{z|x}^{k+1}}^2-\beta\frac{\kappa}{\varepsilon}\norm{B\Delta_* p_{z|x}^{k+1}}^2\\
        &=\left(1-\beta\frac{\kappa}{\varepsilon}\right)\norm{B\Delta_* p_{z|x}^{k+1}}^2.
    \end{split}
\end{equation*}
The proof is complete by defining the scalar $\gamma_\beta:=(\beta\kappa/\varepsilon-1)$.
\end{IEEEproof}
 
 As a remark, \eqref{eq:g_grad} is similar to the weak convexity condition in \cite{zhang_shen_2019}, but different from their method, we do not assume the matrix $B$ to have full column rank.
\begin{lemma}\label{lemma:p_zx_lower_bound}
With the proposed algorithm, we have:
    \begin{multline}\label{eq:lp_2_first_neg_term}
        \norm{B\Delta_*p_{z|x}^{k+1}}^2\leq\mthes{\left({\frac{1}{c^2}+\frac{1-\alpha}{c\alpha}}\right)\norm{\mu_z^k-\mu_z^{k+1}}^2\\
        +\left({1+\frac{1}{c(1-\alpha)}}\right)\norm{\Delta_*p_z^{k+1}}^2},
    \end{multline}
    where $\alpha$ is a scalar and $\alpha\in(0,1)$.
\end{lemma}
\begin{IEEEproof}
From \eqref{eq:min_condition} and recalling the definition $\Delta_* p_{z|x}^{k+1}: p_{z|x}^{k+1}- p_{z|x}^{*}$, we have:
\begin{equation*}
\begin{split}
    \norm{B\Delta_* p_{z|x}^{k+1}}^2&=\mthes{\frac{1}{c^2}\norm{\mu_z^{k}-\mu_z^{k+1}}^2+\norm{\Delta_* p_z^{k+1}}^2\\
    &\quad+\frac{2}{c}\angdot{\mu_z^k- \mu_z^{k+1},\Delta_* p_z^{k+1}}}\\
    &\leq \bigg[\left({\frac{1}{c^2}+\frac{1-\alpha}{c\alpha}}\right)\norm{\mu_z^k-\mu_z^{k+1}}^2\\
    &\quad+\left({1+\frac{1}{c(1-\alpha)}}\right)\norm{\Delta_*p_z^{k+1}}^2\bigg].
    \end{split}
\end{equation*}
 Note that the last inequality follows from $2\langle{u,v\rangle}\leq \frac{1-\alpha}{\alpha}\lVert{u}\rVert^2+\frac{\alpha}{1-\alpha}\lVert{v}\rVert^2, 0<\alpha<1$.
\end{IEEEproof}
The above lemma shows that $\norm{B\Delta_*p_{z|x}^{k+1}}$ is upper bounded by a linear combination of $\norm{\Delta_*p_z^{k+1}}$ and $\norm{\mu_z^{k+1}-\mu_z^k}$. Along with the next lemma, the convergence of the proposed algorithm will be proved by proving that the two sequences $\{\norm{p_z^{k+1}-p_z^k}\},\{\norm{\mu_z^{k+1}-\mu_z^k}\}$ are convergent.
\begin{lemma}\label{lemma:strong_cvx_F}
Suppose that the assumptions stated at the start of Section \ref{parag:assumptions} hold and there exists at least one stationary point $q^*:=(p_z^*,p_{z|x}^*,\mu_z^*)$ satisfying \eqref{eq:sta_pts}. Define
\begin{equation}\label{eq:def_lyapunov_V}
  \tilde{V}^k:=\frac{c}{2}\norm{\Delta_*p_z^k}^2+\frac{1}{2c}\norm{\Delta_*\mu_z^k}^2. 
\end{equation}
Then, if there exists $\alpha\in(0,1)$ such that $\rho_1:=\eta_z-\gamma_\beta[1+\frac{1}{c(1-\alpha)}]$ and $\rho_2:=\frac{1}{2c}-\gamma_\beta(\frac{1}{c^2}+\frac{1-\alpha}{c\alpha})$ are non-negative, $\{\tilde{V}^k\}$ is a non-increasing sequence, namely:
\begin{multline*}
\tilde{V}^k-\tilde{V}^{k+1}\geq \rho_1\norm{\Delta_*p_z^{k+1}}^2+\rho_2\norm{\mu_z^{k+1}-\mu_z^k}^2\\
+\rho_3\norm{p_z^k-p_z^{k+1}}^2,  
\end{multline*}
where $\rho_3:=\eta_z+\frac{c}{2}$.
\end{lemma}
\begin{IEEEproof}
See Appendix \ref{app:pf_of_lyapunov_func}.
\end{IEEEproof}
From Lemma \ref{lemma:strong_cvx_F}, we see that the sequence $\{\tilde{V}^k\}$ is non-increasing if $\{\rho_i\}_{i=1}^3$ are non-negative. The following lemma validates the existence of such condition.
\begin{lemma}\label{cor:1_no_c_involve}
Suppose that the assumptions stated at the start of Section \ref{parag:assumptions} hold, then the condition of Lemma \ref{lemma:strong_cvx_F} is met and the sequence $\{\tilde{V}_k\}$ therein is non-increasing if $\alpha > 1-\frac{1}{2\eta_z}$, where $\alpha\in(0,1)$ is the free parameter in Lemma \ref{lemma:strong_cvx_F}.
\end{lemma}
\begin{IEEEproof}
Using the notation of Lemma \ref{lemma:strong_cvx_F}, we note that $\rho_3$ is always positive. Suppose we let $\rho_1=0$, then
$\eta_z = \gamma_\beta\mthes{1+\frac{1}{c(1-\alpha)}}$.
Now, for $\rho_2$ to be non-negative, we need the relation:$\frac{1}{2}\geq\gamma_\beta\pthes{\frac{1}{c}+\frac{1-\alpha}{\alpha}}$. Putting them together, we have:
\begin{align*}
    \frac{1}{2}&\geq\frac{\eta_z}{1+\frac{1}{c(1-\alpha)}}\left({\frac{1}{c}+\frac{1-\alpha}{\alpha}}\right)\\
    &> \eta_z\pthes{\frac{1-\alpha}{\alpha}}\alpha=\eta_z(1-\alpha),
\end{align*}
where in the last inequality we use $\alpha<\frac{\alpha+c(1-\alpha)}{1+c(1-\alpha)}$.
\end{IEEEproof}
Following lemmas \ref{lemma:strong_cvx_F} and \ref{cor:1_no_c_involve}, proving the convergence of the proposed algorithm would follow from showing that the three sequences $\{\norm{\Delta_*p_z^{k+1}}^2\},\{\norm{p_z^{k+1}-p_z^{k}}^2\},\{\norm{\mu_z^{k+1}-\mu_z^k}^2\}$ are bounded and tend to $(p_z^*,\mu_z^*)$ as $k\rightarrow \infty$.
\begin{theorem}\label{thm:main_thm}
Under the assumptions stated at the start of Section \ref{parag:assumptions}, suppose that there exists a cluster of stationary points of $\mathcal{L}_c(p_z,p_{z|x},\mu_z)$ in \eqref{eq:main_alg}. Namely, \eqref{eq:sta_pts} is satisfied for each.  Denote that cluster as $\Omega$, then with the proposed algorithm, the sequence $q^k:=\{p_z^k,p_{z|x}^k,\mu_z^k\}$ is convergent. Furthermore, $q^k\rightarrow q^*$ as $k\rightarrow \infty$, where $q^*=(p_z^*,p_{z|x}^*,\mu_z^*)\in \Omega$.
\end{theorem}
\begin{IEEEproof}
Recall from lemmas \ref{lemma:strong_cvx_F} and \ref{cor:1_no_c_involve}, we have the non-increasing sequence:
\begin{multline*}
\tilde{V}^k-\tilde{V}^{k+1}\geq \rho_1\norm{\Delta_*p_{z}^k}^2+\rho_2\norm{p_z^{k+1}-p_z^k}^2\\
+\rho_3\norm{\mu_z^{k+1}-\mu_z^k}^2.   
\end{multline*}
Summing both sides from $k=1$ to $k\rightarrow \infty$, we have:
\begin{multline}\label{eq:pf_vdiff_bounded}
\tilde{V}^{1}-\tilde{V}^{\infty}\geq\rho_1\sum_{k=1}^{\infty}\norm{\Delta_*p_z^k}^2+\rho_2\sum_{k=1}^{\infty}\normdiff{p_z^{k+1}}{p_z^{k}}^2\\
+\rho_3\sum_{k=1}^{\infty}\normdiff{\mu_z^{k+1}}{\mu_z^k}^2.    
\end{multline}
Further, we notice from \eqref{eq:def_lyapunov_V} that the non-increasing  $\tilde{V}^k$ is lower-semi-continuous. Hence, $\liminf_{k\rightarrow \infty}\tilde{V}^k=\tilde{V}^*$ and $\tilde{V}^k\leq\tilde{V}^1$, therefore $\tilde{V}^{1}-\tilde{V}^{\infty}$ is bounded.
Since the right hand side of \eqref{eq:pf_vdiff_bounded} must be finite and $\rho_3>0$, $\sum_{k=1}^{\infty}\normdiff{p_z^{k+1}}{p_z^{k}}^2$ is bounded which gives $p_z^k \rightarrow p_z^*$ as $k\rightarrow \infty$. In turn, it immediately follows that $\sum_{k=1}^{\infty}\normdiff{\mu_z^{k+1}}{\mu_z^k}^2$ is bounded and $\normdiff{\mu_z^{k+1}}{\mu_z^k}^2\rightarrow 0$ as $k\rightarrow \infty$. Now, since $\tilde{V}^k\rightarrow 0$, $\mu_z^k\rightarrow \mu_z^*$ as $k\rightarrow \infty$. 
From Lemma \ref{lemma:p_zx_lower_bound}, $\sum_{z}\norm{B\Delta_*p_{z|x}^{k+1}}^2$ is upper bounded by a linear combination of $\sum_{z}\norm{\mu_z^{k+1}-\mu_z^{k}}$ and $\sum_{z}\norm{\Delta_*p_z^{k+1}}$, which are bounded as previously shown. Therefore, similarly, $p_{z|x}^{k+1}\rightarrow p_{z|x}^*$ as $k\rightarrow \infty$. In conclusion, we have shown that $\{p_z^k,p_{z|x}^k,\mu_z^k\}$ is a convergent sequence and is converging toward $(p_z^*,p_{z|x}^*,\mu_z^*)$ as $k\rightarrow\infty$.
\end{IEEEproof}

 \begin{figure*}
   \centerline{
        \subfloat[Information plane]{
           \includegraphics[width=2.5in]{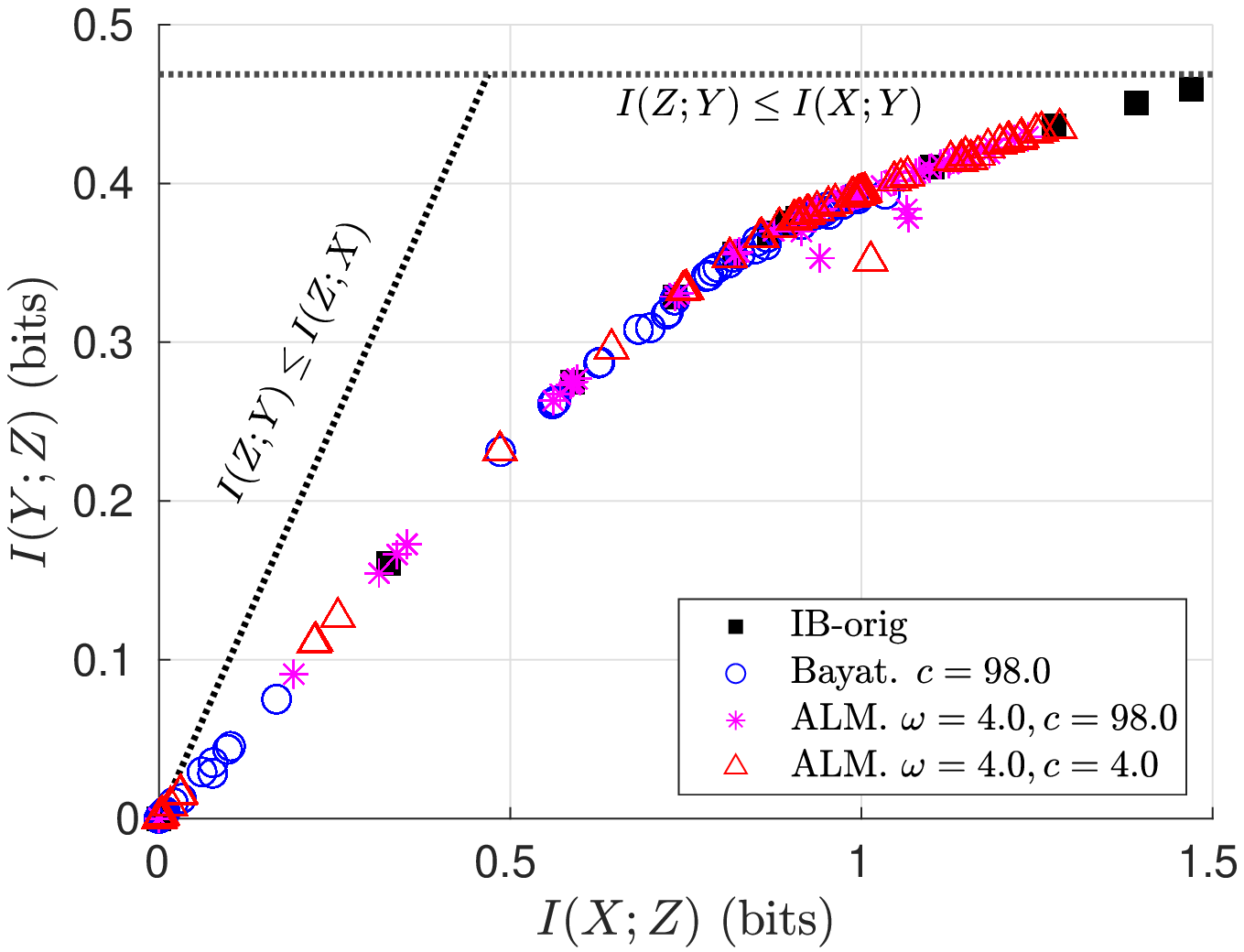}
           \label{subfig:ibcurve}}
        \hfil
        \subfloat[Mutual information $I(Y;Z)$]{
           \includegraphics[width=2.5in]{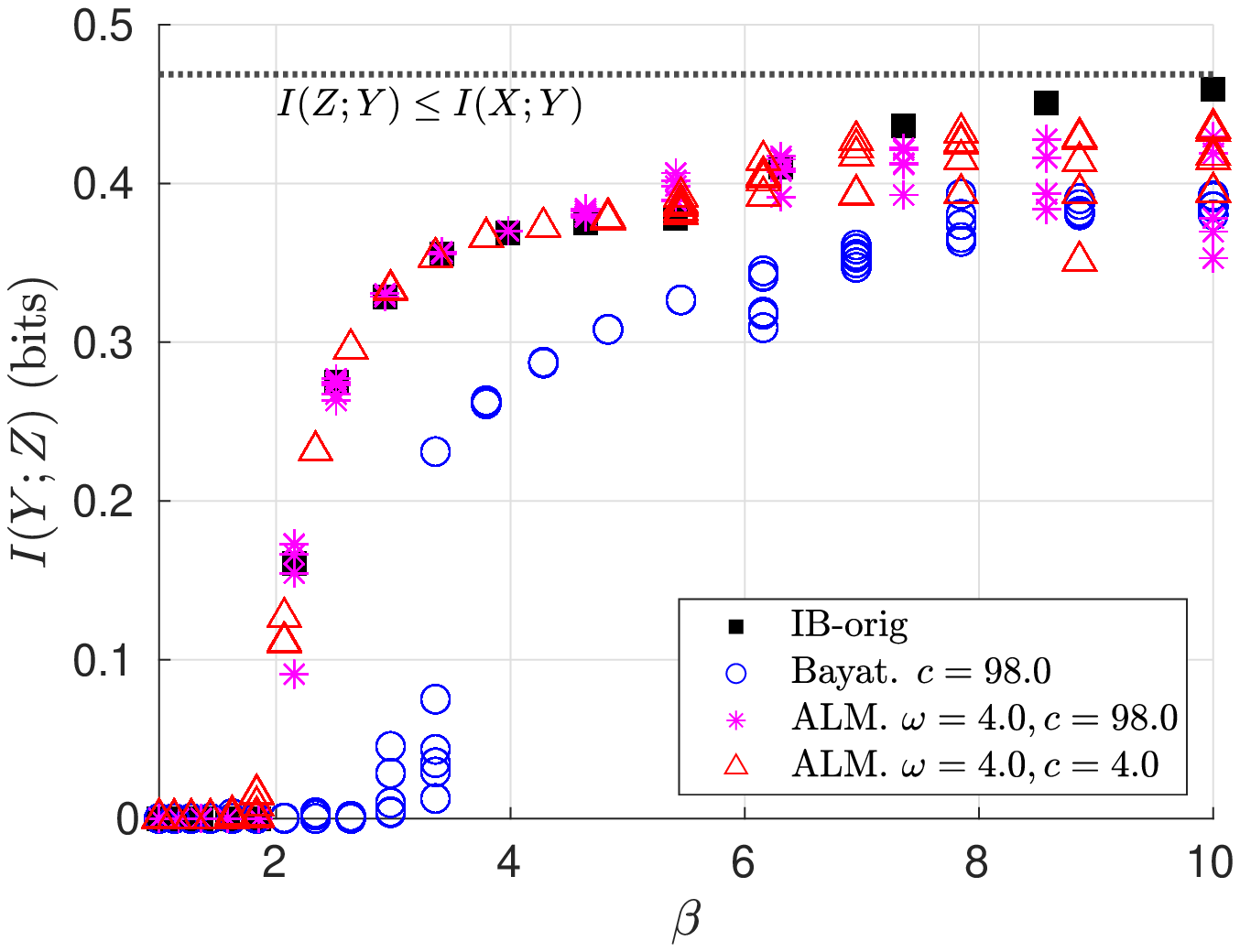}
           \label{subfig:miyz}
        }
   }
   \centerline{
        \subfloat[Percentage of Convergent Cases, $\omega=4.00$]{
            \includegraphics[width=2.5in]{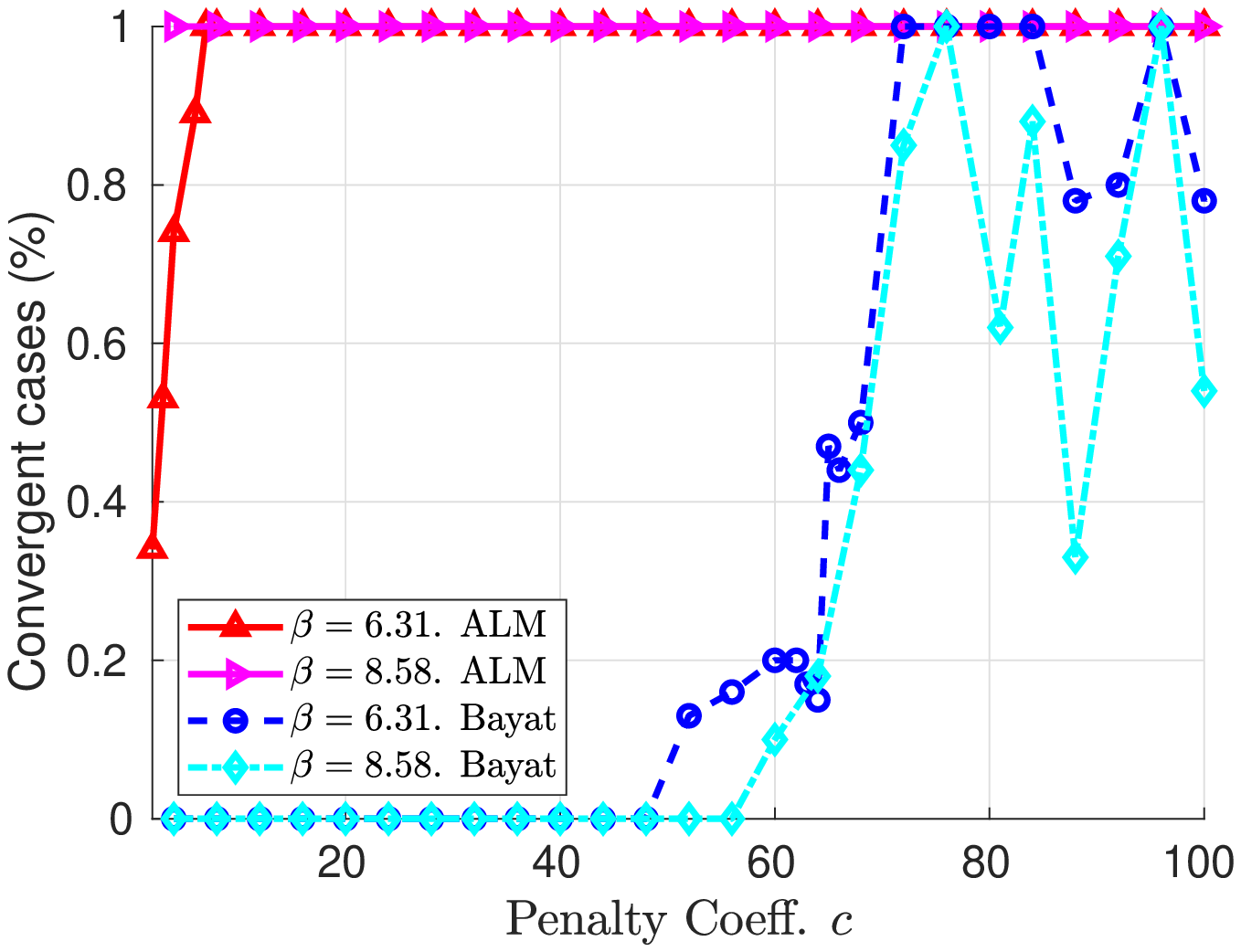}
            \label{subfig:conv_percent}
        }
        \hfil
        \subfloat[Average CPU time per run]{
            \includegraphics[width=2.5in]{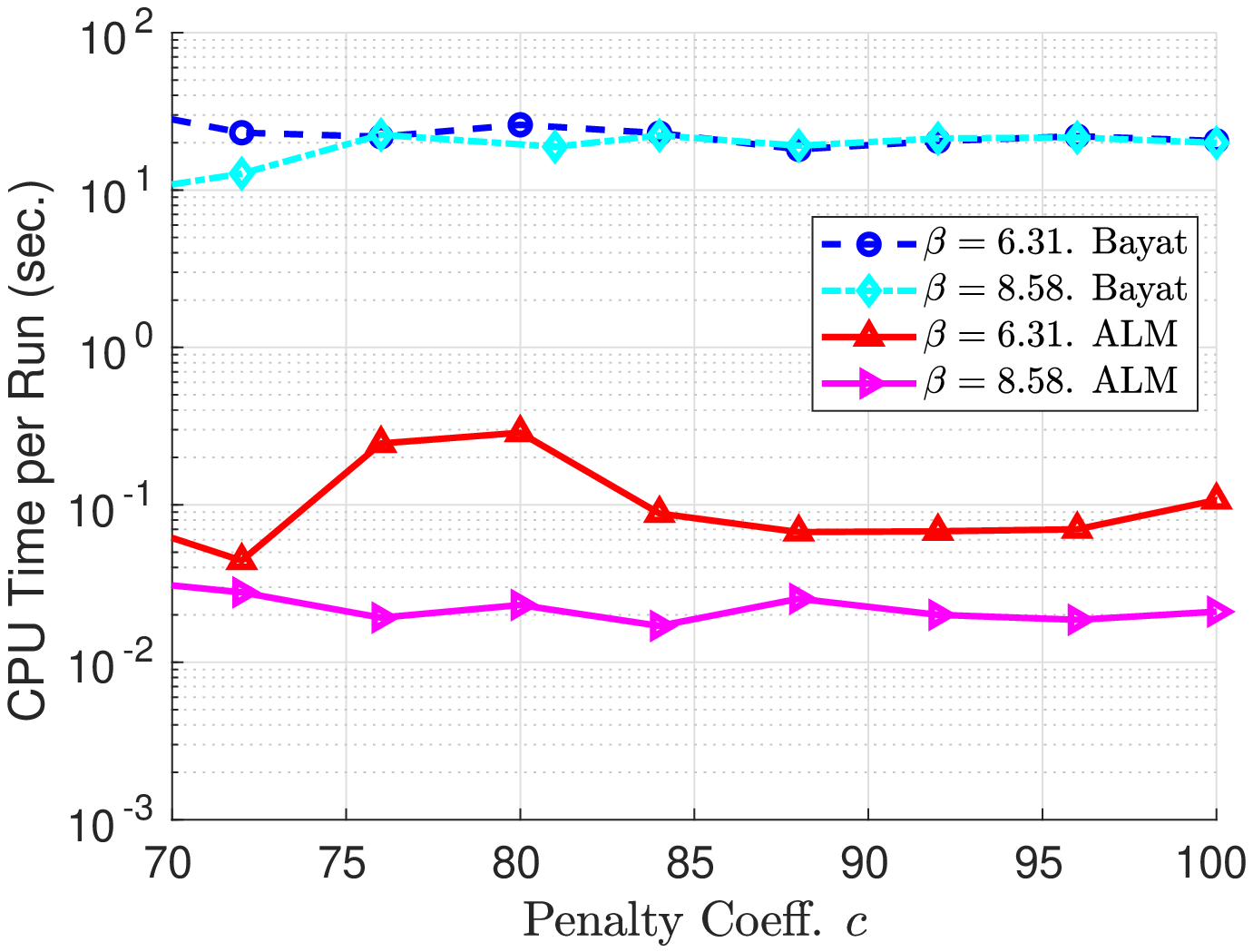}
            \label{subfig:cpu_time}
        }
   }
   \caption{Simulation results on a synthetic joint probability $p(X,Y)$ as in \cite{Wu2020Phase}. In Figure \ref{subfig:ibcurve} and \ref{subfig:miyz} only convergent cases are scattered on the figures. In Figure \ref{subfig:conv_percent} and \ref{subfig:cpu_time}, the regularization parameter is $\omega=4.0$ for the proposed method (ALM). For each method we take average over $100$ runs including both convergent and divergent cases.}
   \label{fig:c3_all}
 \end{figure*}
\section{Evaluation}\label{sec:V_evaluation}

Here, we present simulation results for the proposed algorithm using a synthetic dataset. We implement our method with gradient descent to update the primal variables $p_z,p_{z|x}$:
\begin{equation}\label{eq:update_gd}
    \begin{split}
        p^{k+1}_{z|x}:=&p^{k}_{z|x}-\epsilon^k_{z|x}\nabla_{z|x}\bar{\mathcal{L}}_c^{k},\\
        p^{k+1}_z:=&p^{k}_z-\epsilon^k_z\nabla_z\bar{\mathcal{L}}_c^{k},\\
    \end{split}
\end{equation}
where $\nabla_{z|x}\bar{\mathcal{L}}^k_c:=\nabla_{z|x}\mathcal{L}^k_c-\frac{1}{N_z}\sum_{z}\nabla_{z|x}\mathcal{L}^k_c,\nabla_z\bar{\mathcal{L}}^k_\beta:=\nabla_{z}\mathcal{L}^k_c-\frac{1}{N_z}\sum_{z}\nabla_z \mathcal{L}^k_c$ denote the mean subtracted gradients and $\epsilon_z,\epsilon_{z|x}$ are step sizes for $p_z$, $p_{z|x}$ that assure that the updates remain valid (conditional) probabilities. For comparison, we re-implement \cite{8919752} (denoted as \textit{Bayat}). Note that although there are two versions where one fixed the primal update order the other randomize it, the two methods give comparable performance and we chose the fix-ordered one. We adopt a synthetic joint probability $P(X,Y)$ as in \cite{Wu2020Phase}, where $N_x=N_y=N_z=3$. 
\[
p(Y|X)=\begin{bmatrix}0.7&0.3&0.075\\0.15 & 0.5 & 0.025\\ 0.15& 0.2 & 0.9\end{bmatrix},\quad p(X)=\begin{bmatrix}\frac{1}{3}&\frac{1}{3}&\frac{1}{3}\end{bmatrix}^T.
\]

We evaluate the solutions of the proposed algorithm to conventional methods on the information plane. For each algorithm, we randomly initialize $p_{z|x},p_z$ and run the algorithm until it terminates $100$ times for each  $\beta\in[1.0,10.0]$. For the proposed method, the stopping criterion is either $\norm{p_z^{k}-Bp_{z|x}^{k}}_1^2<2\times 10^{-6}$ (convergent case) or a maximum number of iterations is reached (divergent case). On the other hand, \cite{8919752} imposes an additional set of penalty constraints and therefore an extra condition $\forall z\in\mathcal{Z},\forall y\in\mathcal{Y}$:
\[
\norm{p^{k}(z|y)-\sum_{x}p(z|x)p(x|y)/\sum_{x,z}p(z|x)p(x|y)}<2\times 10^{-6}.
\]

As shown in Figure \ref{subfig:ibcurve}, the solutions obtained through the proposed method (denoted as \textit{ALM}) converge to points comparable to those obtained through the BA-based algorithm (denoted as \textit{IB-orig}), which are considered as the benchmark. Note that for \textit{Bayat}, we have to set a relatively high penalty coefficient $c=98.0$ to ensure convergence compared to ours.

Then we focus on comparing our method to \textit{Bayat}. As shown in Figure \ref{subfig:conv_percent} and \ref{subfig:cpu_time}, our method is convergent for a wider range of the penalty coefficient $c\in[4,100]$. Moreover, when the percentage of convergence cases exceeds $30\%$, the proposed method is convergent with significantly less average CPU time per run (until a termination condition is reached). The results clearly demonstrates the advantage of our two-block design over the three-block ADMM of \cite{8919752}.

\section{Conclusion}\label{sec:VI_conclusion}

In this work, we proposed an ADMM solver for the IB Lagrangian. Unlike previous works, we proved that the proposed algorithm is convergent. Moreover, in proving convergence, we showed that the weak convexity assumption can be relaxed when applying ADMM to the IB objective. Further, empirically, we obtained denser and comparable information plane solutions to the BA-based algorithm. As a final remark, we note that the proposed method links IB problems to the recent global convergence results of ADMM for nonconvex, nonsmooth objectives\cite{zhang_shen_2019,bauschke2017descent}. We hence believe that a rigorous discussion of the conditions for ADMM global convergence in IB problems can lead to further improvement of the convergence guarantee for the proposed framework to global convergence. 
In addition, as the proposed method relies on gradient descent optimization, it opens the door for possible combination with deep neural networks\cite{8550778,sun2016deep} to solve large-scale problems, which is left for future investigation. 


\appendices
\section{Proof of Lemma \ref{lemma:bregman_str_cvx}}\label{app:pf_of_breg_reg}
The goal is to show that the $(\beta-1)$-\textit{strongly convex} function $F_\beta(p_z)$ plus the $\omega$-\textit{strongly convex} function $D_{KL}(p_z^{k+1}||p_z^{*})$ is $(\beta+\omega-1)$-\textit{strongly convex}.
\begin{IEEEproof}
Consider the following
\begin{align*}
        &\angdot{\Delta_*\nabla_z F_\beta^{k+1}+\omega \nabla_z D_{KL}(p_z^{k+1}||p_z^{k}),\Delta_* p_z^{k+1}}\\
        &=\angdot{\Delta_*\nabla F_\beta^{k+1}+\omega\pthes{\nabla \phi_z^{k+1}-\nabla \phi_z^k},\Delta_* p_z^{k+1}}\\
        &=(\beta+\omega-1)\pthes{D_{KL}[p_z^{k+1}||p_z^*]+D_{KL}[p_z^{*}||p_z^{k+1}]}\\
        &\quad-\omega D_{KL}[p_z^*||p_z^k]-\omega\sum_zp_z^{k+1}\log{\left(\frac{p_z^k}{p_z^{k+1}}\frac{p_z^{k+1}}{p_z^*}\right)}\\
        &\begin{multlined}[t]
            =(\beta+\omega-1)\bracket{D_{KL}[p_z^{k+1}||p_z^*]+D_{KL}[p_z^{*}||p_z^{k+1}]}\quad\\
            +\omega\bracket{D_{KL}[p_z^{k+1}||p_z^k]-D_{KL}[p_z^{k+1}||p_z^*]\\
            -D_{KL}[p_z^*||p_z^k]}
        \end{multlined}\\
        &\geq(\beta+\omega-1)\bracket{D_{KL}[p_z^{k+1}||p_z^*]+D_{KL}[p_z^{*}||p_z^{k+1}]}\\
        &\geq(\beta+\omega-1)\Norm{\Delta_*p_z^{k+1}}_1^2\geq (\beta+\omega-1)\Norm{\Delta_*p_z^{k+1}}_2^2 .
\end{align*}
Note that the first two inequalities follow from the Pythagorean theorem \cite{Cover:2006:EIT:1146355} then Pinsker's inequalty. 
For the first inequality to hold, we need to define a convex set that includes $p_z^{k+1},p_z^*$. Let it be $\mathcal{E}_\delta:=\Omega\cup\{q_z:\mathcal{L}_c(q_z)-\mathcal{L}_c(q_z^*)\leq \delta\}$, $\delta>0$. From the convexity of  $\mathcal{F}_\beta(p_z)+\omega D_{KL}(p_z||p_z^{k-1})$ and $\norm{p_z-Bp_{z|x}^{k+1}}$ w.r.t. $p_z$, it follows that $\mathcal{E}_\delta$ is a convex set and $\mathcal{E}_\delta\rightarrow \Omega$ as $\delta\rightarrow 0$.
\end{IEEEproof}

\section{Proof of Lemma \ref{lemma:strong_cvx_F}}\label{app:pf_of_lyapunov_func}
\begin{IEEEproof}
Consider the gradient of $F_\beta(p_z)$ following the minimizer condition \eqref{eq:min_condition}:
\begin{multline*}
\nabla_z\mathcal{L}_c^k\pthes{p_z^{k+1},p_{z|x}^{k+1},\mu_z^k}=0\\
\Rightarrow \nabla_zF_\beta^{k+1}+\omega\left[\nabla_z\phi(p_z^{k+1})-\nabla_z\phi(p_z^k)\right]=-\mu_z^{k+1}.
\end{multline*}
Similarly, the gradient of $\mathcal{L}_c(p_z,p_{z|x},\mu_z)$ w.r.t. $p_{z|x}$ is:
\begin{equation*}
\nabla_{z|x}G_\beta^{k+1}=B^T\mthes{\mu_z^{k+1}-c\pthes{p_z^{k+1}-p_z^k}}.
\end{equation*}
Hence, we have:
\begin{equation}\label{eq:pv_v_lyp_form}
\begin{split}
    &\begin{multlined}[t]
        \angdot{{\Delta_*\nabla_zF_\beta^{k+1}+\omega\left[\nabla_z\phi(p_z^{k+1})-\nabla_z\phi(p_z^{k})\right]},\Delta_*p_z^{k+1}}\\
        \quad+\angdot{\Delta_*\nabla_{z|x}G_\beta^{k+1},\Delta_*p_{z|x}^{k+1}}
    \end{multlined}\\
    =&\begin{multlined}[t]
    -\angdot{\Delta_*\mu_z^{k+1},\Delta_* p_z^{k+1}}+\angdot{\Delta_* \mu_z^{k+1},B\Delta_* p_{z|x}^{k+1}}\\
    -c\angdot{\Delta_* p_z^{k+1},B\Delta_* p_{z|x}^{k+1}}+c\angdot{\Delta_* p_z^{k},B\pthes{\Delta_* p_{z|x}^{k+1}}}
    \end{multlined}\\
    =&\begin{multlined}[t]
        -\frac{1}{c}\angdot{\Delta_*\mu_z^{k+1},\Delta_*\mu_z^{k+1}-\Delta_*\mu_z^{k}}\\
        -c\angdot{\Delta_* p_z^{k+1},B\Delta_* p_{z|x}^{k+1}}+c\angdot{\Delta_* p_z^{k},B\Delta_* p_{z|x}^{k+1}}
    \end{multlined}\\
    =&\begin{multlined}[t]
        \frac{1}{2c}\left(\norm{\Delta\mu_z^k}^2-\norm{\Delta\mu_z^{k+1}}^2\right)-\frac{c}{2}\norm{\Delta p_z^{k+1}}^2\\+\frac{c}{2}\left(\norm{\Delta p_z^k}^2
        -\norm{\Delta p_z^k-B\Delta p_{z|x}^{k+1}}^2\right),
    \end{multlined}
\end{split}
\end{equation}
where we use the identity:
\[
2\angdot{u-v,w-u}=\normdiff{v}{w}^2-\normdiff{u}{v}^2-\normdiff{u}{w}^2.
\]
Then recall the Lyapunov function defined in \eqref{eq:def_lyapunov_V}:
\begin{equation*}
  \tilde{V}^k:=\frac{c}{2}\norm{\Delta_* p_z^k}^2+\frac{1}{2c}\norm{\Delta_*\mu_z^k}^2.  
\end{equation*}
We can rewrite (\ref{eq:pv_v_lyp_form}) as:
\begin{equation*}
\tilde{V}^{k}-\tilde{V}^{k+1}-\frac{c}{2}\norm{\Delta_* p_z^k-B\Delta_* p_{z|x}^{k+1}}^2.
\end{equation*}
Now, using the $\eta_z$-\textit{strong convexity} of $F_\beta(p_z)$ and Lemma \ref{lemma:gzx_coeff}:
\begin{multline}\label{eq:Lp_two_terms}
    \tilde{V}^{k}-\tilde{V}^{k+1}\geq \eta_z\norm{\Delta_* p_z^{k+1}}^2-\gamma_{\beta}\norm{B\Delta_* p_{z|x}^{k+1}}^2\\
    +\frac{c}{2}\norm{\Delta_* p_z^{k}-B\Delta_* p_{z|x}^{k+1}}^2.
\end{multline}
The next step is to replace the intermediate primal variables $p_{z|x}$ with either the other primal $p_z$ or the dual variables $\mu_z$. For $\norm{B\Delta_*p_{z|x}^{k+1}}^2$, we use Lemma \ref{lemma:p_zx_lower_bound}. For the last term, we have:
\begin{equation}\label{eq:lp_2_second_pos_term}
\begin{split}
    {}&\norm{\Delta_* p_z^k-B\Delta_* p_{z|x}^{k+1}}^2\\
    =&\norm{\Delta_* p_z^k-\Delta_* p_z^{k+1}+\Delta_* p_z^{k+1}-B\Delta_* p_{z|x}^{k+1}}^2\\
    =&\norm{\Delta_* p_z^{k}-\Delta_* p_z^{k+1}+\frac{1}{c}\pthes{\Delta_* \mu_z^{k+1}-\mu_z^{k}}}^2\\
    =&\begin{multlined}[t]
        \norm{\argdiff{p_z}{k}{k+1}}^2+\frac{1}{c^2}\norm{\argdiff{\mu_z}{k+1}{k}}^2\\
        +\frac{2}{c}\angdot{\argdiff{p_z}{k}{k+1},\argdiff{\mu_z}{k+1}{k}}.
    \end{multlined}
    \end{split}
\end{equation}
Plugging \eqref{eq:lp_2_first_neg_term} and \eqref{eq:lp_2_second_pos_term} into \eqref{eq:Lp_two_terms}, we have:
\begin{equation*}
    \begin{split}
        {}&\tilde{V}^k-\tilde{V}^{k+1}\\
        \geq&\begin{multlined}[t]
            \eta_z\norm{p_z^{k+1}-p_z^*}^2-\gamma_\beta\Big\{\pthes{\frac{1}{c^2}+\frac{1-\alpha}{c\alpha}}\norm{\mu^k-\mu^{k+1}}^2\\
            +\mthes{1+\frac{1}{c(1-\alpha)}}\norm{p_z^{k+1}-p_z^*}^2\Big\}+\frac{c}{2}\Big\{
            \norm{\argdiff{p_z}{k}{k+1}}^2\\
            +\frac{1}{c^2}\norm{\argdiff{\mu}{k+1}{k}}^2+\frac{2}{c}\angdot{\argdiff{p_z}{k}{k+1},\argdiff{\mu}{k+1}{k}}
            \Big\}
        \end{multlined}\\
        =& \begin{multlined}[t]
                \bracket{\eta_z-\gamma_\beta\mthes{1+\frac{1}{c(1-\alpha)}}}\Norm{p_z^{k+1}-p_z^*}^2\\
                +\mthes{\frac{1}{2c}-\gamma_\beta\pthes{\frac{1}{c^2}+\frac{1-\alpha}{c\alpha}}}\Norm{\mu^{k+1}-\mu^k}^2\\
                +\frac{c}{2}\norm{p_z^k-p_z^{k+1}}^2+\angdot{p_z^k-p_z^{k+1},\mu^{k+1}-\mu^k}
            \end{multlined}\\
        \geq&
        \begin{multlined}[t]
            \bracket{\eta_z-\gamma_\beta\mthes{1+\frac{1}{c(1-\alpha)}}}\Norm{p_z^{k+1}-p_z^*}^2\\
            +\mthes{\frac{1}{2c}-\gamma_\beta\pthes{\frac{1}{c^2}+\frac{1-\alpha}{c\alpha}}}\Norm{\mu^{k+1}-\mu^k}^2\\
            +\pthes{\eta_z+\frac{c}{2}}\norm{p_z^k-p_z^{k+1}}^2,
        \end{multlined}
    \end{split}
\end{equation*}
where in the last inequality we use the $\eta_z$-strong convexity of $F_\beta(p_z)$ and $\alpha\in(0,1)$. Define $\rho_1:=
        \eta_z-\gamma_\beta[1+\frac{1}{c(1-\alpha)}]
        $, $\rho_2:=
            \frac{1}{2c}-\gamma_\beta(\frac{1}{c^2}+\frac{1-\alpha}{c\alpha})
        $ and $\rho_3:=\eta_z+\frac{c}{2}$. Since $c>0$ and $\eta_z>0$, we always have $\rho_3>0$. Therefore, we only need $\rho_1$ and $\rho_2$ to be non-negative to assure $\tilde{V}^{k}-\tilde{V}^{k+1}$ is non-increasing as $k\rightarrow\infty$.
\end{IEEEproof}



\bibliographystyle{IEEEtran}
\bibliography{definitions,bibliofile}


\end{document}